\documentstyle{basi}
\begin{document}
\title[Asymmetry between velocity and intensity power]{On the p-mode asymmetry between velocity
and intensity from the GONG+ data } 
\author[S. C. Tripathy, Kiran Jain, Frank Hill and C. G. Toner]%
       {S. C. Tripathy$^{1,2}$\thanks{e-mail:sushant@prl.ernet.in}, Kiran Jain$^3$, 
Frank Hill$^4 $ and C. G. Toner$^4$\\ 
 $^1$Department of Physics, Indian Institute of Science, Bangalore 560 012, India \\      
$^2$Udaipur Solar Observatory, PRL, PO Box 198, Udaipur, 313 001, India\\
$^3$Indian Institute of Astrophysics, Koramangala, Bangalore 560 034, India\\
$^4$National Solar Observatory, Tucson, AZ85726-6732, USA}
\maketitle
\label{firstpage}
\begin{abstract}
 We have analyzed the local acoustic spectra of small regions over the solar
surface at different locations from disk center to limb via the technique of ring
diagrams. It is found that the frequency shifts between velocity and
intensity is a function of location on the disk and is higher near the disk
center than those near the limb. 
\end{abstract}


The peaks of solar oscillation {\it p}-modes
observed in velocity and intensity spectral lines are asymmetric (Duvall et
al. 1993). Moreover, this asymmetry is reversed; velocity (V) has negative
asymmetry (more power on the low frequency side) while intensity (I) power
spectrum has positive symmetry (more power on the high frequency side). The
asymmetry is a strong function of frequency and varies weakly with angular
degree. Roxburgh and Vorontsov (1997) have suggested that the reversal in
asymmetry occurs in velocity while the observations suggest that the
reversal occurs in intensity (Nigam et al., 1998). In a phenomenological
model, Nigam et al. (1998) proposed that the reversal is due to the
correlated background noise whose level depends on the characteristic
granulation. 

Recently Georgobiani et al. (2002) have proposed that the asymmetry
reversal is produced by radiative transfer effects and not by correlated
noise. In addition, a recent study of center to limb variation (CLV) of
solar granulation reports that the granulation contrast increases near the
limb (Sanchez Cuberes et al., 2003). Since, the physics of the correlated
noise in not yet fully understood, it is desirable to test these ideas using
a different data source such as GONG+ which now provides high resolution
Dopplergrams of the Sun at each minute interval.

The data used here consist of velocity and intensity images obtained
by the prototype GONG+ instrument at Tucson for the period June 11-13 and
15, 2000 and GONG+ instrument at BBSO during May 25-27 and 29, 2001. The
disk center (-7.5$^0$ to 7.5$^0$) and four areas in longitude from 15$^0$ to 60$^0$
(at an interval of 15 degrees) each of 15 x 15 degrees are extracted and
tracked and remapped on to a grid of 128 x 128 pixels. A 3D FFT (Hill, 1988)
was used to obtain the power as a function of ($\nu $, $k_x$, $k_y$). 
The power spectrum of each individual day are then combined to
enhance the signal. 

\input epsf
\begin{figure}
\begin{center}
\leavevmode
\epsfysize=9.5cm  \epsfbox{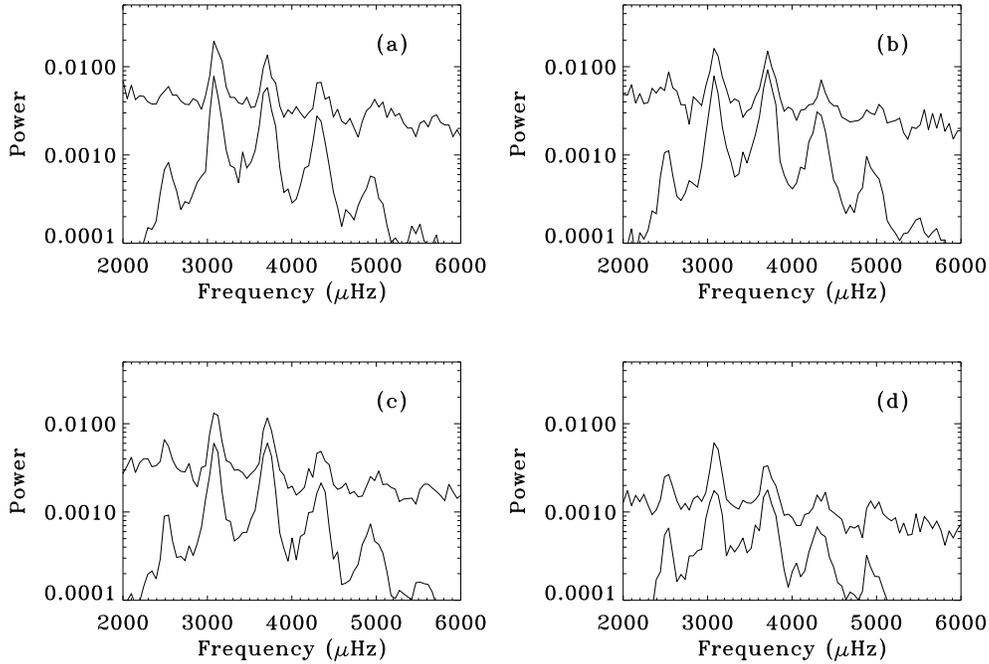}\\
\caption{The combined intensity (top) 
and velocity (bottom) power spectra as a function of center to limb at $\ell $ = 576 
for the GONG+ data from BBSO station. The panels 
 denote the following starting longitudes: (a) -7.5$^0$, (b) 15$^0$, (c) 30$^0$ and (d) 45$^0$.
}  
\end{center}
\end{figure}

Since the numerical magnitude of the measured power in V and I
differ by some 10 orders of magnitude, the spectrum were rescaled to the
interval [0,1] using a range normalization. We further smooth the
renormalized spectra by a boxcar average by 3 bins in $\nu $ and 2 bins each
in $k_{x}$ and $k_{y}$.  
Figure 1 shows the velocity and intensity spectra as a function of
frequency for different regions across CLV. We note the following

\begin{itemize}
\item {At low degrees, the asymmetry agrees with the earlier studies, velocity 
has more power on the low frequency side and intensity has more power on the higher frequency side.}
\item {A frequency shift in V and I is seen above the cutoff frequency of 5.3 mHz 
which is consistent with those of Jain et al. (2003). Also, the frequency shift 
is larger in intensity than velocity. }
\item {The power amplitude is higher near the disk center and decreases towards the limb.}
\item {The frequency shift between V and I at higher frequencies appears to decrease from 
disk center to limb. }
\end{itemize} 

Since the apparent frequency shift between an oscillation observed in velocity and intensity 
cannot be a property of the mode, it must arise from the excitation mechanism. In addition, 
the observed decrease of the V-I frequency shift from center to limb strongly suggests that 
the mechanism is
that of correlated noise from the granulation rather than radiative transfer effect. 
At disk center we observe the vertical velocity field, which has both oscillatory and granulation
contributions. Near the limb we see primarily the horizontal velovity field which is 
dominatd by granulation. This decreases the correlation wih the oscillations and thus
apparently reduces the frequency shift.

We plan to improve the analysis to include additional data, and to quantify the result. The
results will be used to test models of the mode excitation and damping mechanism.

\section*{Acknowledgements}
This work utilises data obtained by 
the Global Oscillation Network Group (GONG) project, managed by the
National Solar Observatory, which is operated by AURA, Inc., under a cooperative agreement with the National
Science Foundation. The data were acquired by instruments operated by the Big Bear Solar Observatory and National 
Solar Observatory at Tucson.

\label{lastpage}
\end{document}